\documentstyle[mncite,psfig]{mn}

\def\et{\em et al. \em}
\def\msun{M_{\odot}}
\def\gte{\,\lower.6ex\hbox{$\buildrel >\over \sim$} \, }
\def\lte{\,\lower.6ex\hbox{$\buildrel <\over \sim$} \, }

\title{The initial conditions of young globular clusters in the LMC}

\author[S.P.Goodwin]{Simon P. Goodwin\\
            Astronomy Centre, University of Sussex, 
            Falmer, Brighton BN1 9QH, UK.}

\date{}

\begin{document}

\maketitle

\begin{abstract}

$N$-body simulations are used to model the early evolution of globular 
clusters.  These simulations include residual gas which was not turned into 
stars which is expelled from the globular cluster by the actions of 
massive ($>8\msun$) stars.  The results of these simulations are compared 
to observations of 8 LMC globular clusters less than 100 Myr old.  These 
observations are used to constrain the initial conditions that may have 
produced these clusters.  
It is found that the observations can be accounted for in a model where 
the globular clusters form from proto-cluster clouds similar to Plummer 
models with length scales in the range $1 < R_{\rm S}/$pc$ < 3$ where 
the star formation efficiency varies between 25\% and 60\%. Using these 
derived initial conditions the survivability of these clusters in both 
the Galaxy and the LMC is assessed.  If the slope of the initial mass 
function is around $\alpha = 2.35$ then 2 or 3 of these clusters may be 
able to survive for a Hubble time even in the Galactic halo.  In this case 
these clusters may represent young versions of the Galactic globular 
cluster population which was severely depleted by the destruction of many 
of its original members.  In the case where $\alpha = 1.50$, however, 
none of these clusters would be expected to survive for more than a few Gyr 
at most, even within the LMC.

\end{abstract}


\section{Introduction}

In this paper the structure of young globular clusters in the Large 
Magellanic Cloud (LMC) is investigated with respect to theoretical models 
of globular cluster formation and evolution.  Observations are compared 
to $N$-body simulations of globular clusters including stellar evolution and 
the expulsion of residual gas not consumed in star formation. This allows 
some limits to be placed on the initial conditions which gave rise to these 
globular clusters.

An important question in the theory of globular cluster formation and 
evolution is how similar are these young LMC objects to the precursors of 
the old globular clusters observed in the Galaxy?  Will these young globular 
clusters also be able to survive for a Hubble time?  This 
paper examines the structure of the young globular clusters found in the 
LMC and tries to determine their initial conditions and, from the present 
theoretical understanding of globular clusters, if they will survive for 
a significant length of time in either the LMC or in the halo of the Galaxy.

Globular clusters are usually extremely chemically homogeneous, compact 
star clusters containing some $10^4$ to $10^6 \msun$ of stars formed, 
presumably, in one burst of star formation.  After this burst of star 
formation it is assumed that any significant amounts of residual gas 
remaining in the globular cluster are expelled.  This prevents further 
star formation in a chemically enriched environment and retains the chemical 
homogeneity of the globular cluster (Lin \& Murray 1991).

The LMC provides the nearest example of the present day formation of globular 
cluster-type objects.  These young LMC objects are variously refered to 
in the literature as young globular clusters, young populous clusters and 
blue globular clusters.  For the purposes of this paper these objects 
will be refered to simply as young globular clusters.

If these young globular clusters are, indeed, similar to those which gave 
rise to the (old) Galactic globular cluster population then their study 
could provide important clues about the initial conditions and extent of 
the Galactic globular clusters.  The closeness of the LMC allows these 
objects to be studied in detail and their structural parameters to be 
determined with some accuracy.  Observations reveal a number of 
general characteristics of young globular clusters in the LMC: 

\begin{enumerate}

\item The density profiles in the inner regions of even very young clusters 
appear to be relaxed and well described by King (1966) models (Chrysovergis, 
Kontizas \& Kontizas 1989 and references therein).

\item  Spatial density profiles fall off in the outer regions of these  
clusters as a power law with index $\gamma + 1 \approx 3.5$ where $\gamma$ is 
a fitting parameter used by Elson, Fall \& Freeman (1987, hereafter EFF) 
defined in section 2.1.

\item Clusters are often found to have large, unbound stellar halos which 
may contain up to 50\% of the mass of the cluster (EFF, van den Bergh 1991).

\end{enumerate} 

\noindent It is a young globular cluster with these characteristics that 
this paper attempts to model with $N$-body simulations.

$N$-body simulations have been recently used to model globular clusters, 
Fukushige \& Heggie (1995) used a $N$-body code to model globular clusters 
and compared these simulations to the Fokker-Planck calculations of 
Chernoff \& Weinberg (1990).  In this case the results of both simulations 
were found to be qualitatively similar.  Such 
results show that the use of $N$-body codes for the modelling of systems 
as large as globular clusters is not without justification.  However, as 
pointed out by Heggie (1995) $N$-body simulations are only of use as 
statistical tests of a system's behaviour.  For this reason, the evolution 
of any one cluster must be tested over a statistically significant number 
of simulations before any robust conclusions can be drawn about its behaviour.

The effects of the expulsion of residual gas from an $N$-body system (an 
open cluster) were first studied by Lada, Margulis \& Dearborn (1984) 
who found that the expulsion of over 50\% of the mass of a cluster may 
not entirely disrupt that cluster, but can leave a bound core of stars.  
The inclusion of residual gas and the effects of its expulsion in  
$N$-body simulations of globular clusters has been investigated in a paper 
by Goodwin (1996, hereafter paper I) who finds that, while the 
expulsion of a large fraction of a globular clusters initial mass is 
disruptive to the cluster, many clusters may well be able to survive in the 
Galaxy with star formation efficiencies (SFEs) as low as 20\% to 25\% 
with sufficiently strict initial conditions. 

This paper assumes that the formation mechanism of globular clusters forms 
a smooth, relaxed initial stellar distribution similar to the 
observed distributions of young globular clusters only $\approx 20$ Myr 
old (Elson, Fall \& Freeman 1989; Elson 1991).  In practice the 
distribution must be relaxed and smooth before the expulsion of the 
cluster's residual gas, not necesserally initially.  It is assumed that 
any substructure and clumpiness present in the initial distribution does 
not significantly effect the dynamics and has been erased by this point.

The survivability of these LMC clusters in the environment of our Galaxy is 
assessed by comparing their postulated initial conditions with theoretical 
calculations of the initial conditions required in the Galaxy.  The 
theoretical constraints used are those from paper I based upon the 
King model based simulations of Chernoff \& Shapiro (1987).  Chernoff \& 
Shapiro presented calculations of the minimum King model required to survive 
for a Hubble time in the Galaxy for a variety of initial masses, initial mass 
function slopes and Galactocentric radii.  These results are found to be in 
reasonable agreement with both Fokker-Planck (Chernoff \& Weinberg 1990) and 
$N$-body (Fukushige \& Heggie 1995) calculations.  These constraints 
were extended in 
paper I to include reasonable star formation rates and the expulsion 
of the residual gas which was not used in star formation.  These results 
provided a minimum scale length of a Plummer model required for survival 
for the range of initial conditions in Chernoff \& Shapiro, star 
formation efficiency and the mechanism for gas expulsion.

\section{Observations}

Eight young LMC globular clusters with well-determined parameters and 
ages less than 100 Myr have been chosen as indicative of the young clusters 
in the LMC.  These clusters are NGC 1818, NGC 2004, NGC 2156, NGC 2157, 
NGC 2159, NGC 2164, NGC 2172 and NGC 2214. Their structural parameters and 
ages are summarised in table 1.  These clusters were all studied by EFF and 6 
of them by Chrysovergis \et (1989).  EFF also included NGCs 
1831 and 1866 in their original sample.  These clusters have not been 
included in the main analysis as they are both older than 100 Myr. 

\begin{center}
\begin{table}
\begin{center}
\begin{tabular}{|c|c|c|c|c|} \hline
NGC & log $M$ & log $\rho_{0}$ & $r_{\rm c}$ & $\tau$ \\ 
  &  $(\msun)$ & $(\msun$ pc$^{-3}$) & (pc) & (Myr) \\ 
  &  (1) & (2) & (3) & (4)   \\ \hline
1818 & 4.69 & 1.6-3.2 & 2.1 & 30 \\ 
2004 & 4.60 & 2.0-3.7 & 1.1 & 20 \\ 
2156 & 4.47 & 2.0-3.2 & 1.7 & 65 \\ 
2157 & 4.60 & 2.0-3.4 & 2.3 & 32 \\ 
2159 & 5.00 & 1.7-2.9 & 1.9 & 65 \\ 
2164 & 4.69 & 2.1-3.4 & 1.5 & 98 \\ 
2172 & 4.40 & 1.5-2.7 & 2.1 & 65 \\ 
2214 & 4.30 & 1.4-2.7 & 2.3 & 83 \\ \hline
\end{tabular}
\end{center}
\caption{Structural parameters of the 8 young LMC globular clusters used 
as comparisons. \protect\newline 
Col. (1) Mass.  Taken from Chrysovergis \et (1989), except NGC 
2159 and NGC 2172 which are the mean of the values given by EFF. 
\protect\newline 
Col. (2) Central density.  From EFF. \protect\newline 
Col. (3) Core radius.  From Elson, Freeman \& Lauer (1989). \protect\newline  
Col. (4) Age.  NGC 1818 average from Will, Bomans \& de Boer (1995).  NGCs  
2004, 2164, 2214 from Girardi \et (1995).  Others in EFF from Hodge (1983). }
\end{table}
\end{center}

It should be noted that the study of Chrysovergis \et (1989) find, 
sometimes substantially, different core radii to Elson, Freeman \& Lauer 
(1989).  These differences are as high as a factor of 2.  The values of 
Elson \et (1989) have been used as they are the values used in 
the profile fittings of EFF.

The quoted masses for these globular clusters should be considered as 
lower limits on the possible mass.  EFF quote the masses over a large 
range (sometimes over an order of magnitude).  Chysovergis \et (1989) 
find masses that are usually at the lower limit of those quoted by EFF, but it 
should be noted that Lupton \et (1989) find masses for NGC 2164 and NGC 2214 
of $2 \times 10^5 \msun$ and $4 \times 10^5 \msun$ respectively.

Of particular interest in this sample are the `quartet' of globular clusters 
NGCs 2156, 2159, 2164 and 2172.  The quartet clusters are located together 
and 3 of them (NGCs 2156, 2157 and 2172) are coeval.  The profiles of these 
clusters are, however, very different.

A question may be raised over the inclusion of NGC 2214 within this 
study.  There is evidence that NGC 2214 may be the result of a merger, 
or may even be in the process of merging now (Bhatia \& MacGillivray 1988).  
There is argument over the possible existence of two different age 
subgiant branches (Sagar, Richtler \& de Boer 1991 and Lee 1992) and 
so NGC 2214 has been kept in the sample.

\subsection{Cluster profiles}

EFF constructed density profiles of the 8 clusters in this sample.  These 
profiles were of the form (equation (13a) in EFF)

\begin{equation}
\rho(r) = \rho_{0} \left( 1 + \frac{r^2}{a^2} \right) ^{-(\gamma + 1)/2}
\end{equation}

where $a$ is a characteristic radius given by 

\[
a = r_{\rm c} (2^{2/\gamma} - 1)^{-1/2}
\]

The density profiles were formed by a deprojection of the luminosity 
profiles.  This deprojection to a density profile assumed a mass-to-light 
ratio and that it is constant over the whole cluster.

These density profiles are very similar to King (1966) models at low 
radii (the radii containing approximately 70\% of the mass of the cluster 
and less), while at higher radii they fall off below a King profile.

Values of $\gamma$ and $a$ for the 8 clusters are shown in table 2.  Using 
these values and those from table 1, it is possible to create the density 
profiles of the clusters.  Integrating this in each case then gives the 
mass distribution.  The mass distribution is prefered as it is easier to 
compare with the $N$-body models as the relatively low number of particles 
used in the simulations make the construction of a smooth density profile 
difficult, while the mass distribution is trivially formed from the positions 
of the particles.

\begin{center}
\begin{table}
\begin{center}
\begin{tabular}{|c|c|c|} \hline
NGC & $\gamma$ & $a$ (pc)   \\  \hline
1818 & $2.45 \pm 0.25 $ & 2.4 \\ 
2004 & $2.20 \pm 0.20 $ & 1.2 \\ 
2156 & $2.75 \pm 0.45 $ & 2.1 \\ 
2157 & $2.90 \pm 0.27 $ & 2.9 \\ 
2159 & $2.15 \pm 0.34 $ & 2.0 \\ 
2164 & $2.80 \pm 0.30 $ & 1.9 \\ 
2172 & $3.20 \pm 0.50 $ & 2.9 \\ 
2214 & $2.40 \pm 0.24 $ & 2.6 \\ \hline
\end{tabular}
\end{center}
\caption{The values of $\gamma$ and $a$ used in the density profiles 
from equation 1.  $a$ is calculated from the values of $r_{\rm c}$ in table 1.}
\end{table}
\end{center}

These profiles often show irregularities that are not present in the profiles 
of older clusters (Elson 1991).  EFF suggest that dips and peaks in the 
outer regions of the luminosity profiles of clusters are probably produced 
by the presence of `clumps' of stars.  The presence of these clumps is 
smoothed away in the fitting of the best fit line to the luminosity profile 
and so these clumps are not present in either the density profile or mass 
distributions derived from the luminosity profiles.  As these dips and peaks 
are found in the outer regions of the clusters they do not contain any 
significant fraction of the mass of the cluster and so may not be very 
important for  the limits of the simulations presented in this paper. However, 
they do make the form of the profiles at large radii uncertain and, possibly, 
misleading.

\subsection{Correlations of cluster properties}

A simple statistical analysis of the relationships between various cluster 
properties has been made to search for clues as to the initial conditions 
and formation mechanisms of the young globular clusters.  Application 
of the sample correlation coefficient to the data shows that it is 
consistent with no correlation between $\gamma$ and any other quantity.  
Age is also totally uncorrelated with any other parameter of the clusters.  
Thus the cluster profiles do not appear to represent any sort of 
evolutionary sequence.  This lack of correlations between profile shape 
and other quantities shows that, in these young globular clusters, the 
profiles are determined by some other mechanism(s) that is not dependent 
upon the structural parameters measured now.

\subsection{Other relevant observations}

Observations of LMC globular clusters show that they are more highly 
elliptical than Galactic globular clusters (van den Bergh \& Morbey 1984).  
The reasons for this high ellipticity are unknown.  It is possible that 
LMC clusters have high rotational velocities.  Elson (1991), however, 
suggests that the presence of subclumps merging with the main cluster 
may produce the impression of high ellipticities in these young globular 
clusters.

It has been noted that the half-mass radii $r_{\rm h}$ of globular clusters 
in the LMC are usually 3 to 4 times larger than similar Galactic clusters 
(van den Bergh 1991).  This result is only relevant for the older LMC globular 
clusters as the Galaxy has no analogues of the LMC's young clusters.  The 
half-mass radii from simulations of these clusters can, however, be 
compared to the theoretical constraints on young Galactic globular clusters 
from simulations.

The mass functions of these clusters have been studied by a number of 
authors.  There is some considerable discrepancy about the value of the 
mass function slope in the young clusters in the LMC.  Mateo (1988) finds 
in a study of 6 young and intermediate age LMC and SMC clusters that 
the mass function slopes are consistent with all being drawn from a single 
IMF of slope $\alpha = 2.52$.  Sagar \& Richtler (1991) find slightly lower, 
but still Salpeter-type slopes consistent with $\alpha = 2.1$.  Elson, Fall 
and Freeman (1989), however, in a sample 
of 6 of the young LMC clusters from EFF find that $0.8 < \alpha < 1.8$.  
None of the clusters in these two studies overlap. A recent 
determination of the mass function in NGC 2214 by Banks, Dodd \& 
Sullivan (1995) finds $\alpha \approx 2$ in contrast to  Elson, Fall and 
Freeman's (1989) value of 0.8.  Such discrepancies arise due to  
observational uncertainties and the application of different methods 
for the determination of mass function slopes.  However, it shows 
that little importance should be placed on any individual value.

NGC 2070 is of interest to this study as it is the only LMC globular cluster 
that has not, as yet, expelled its residual gas.  NGC 2070 is the central 
cluster of the 30 Dor nebula with an age of only a few Myr (Meylan 
1993).  NGC 2070 is composed of large clumps of stars with little symmetry 
apparent in the distribution of stars.  This clumpiness in the light 
distribution could, however, be caused by concentrations of high mass 
stars and may not trace the mass distribution.  King model fits to the 
luminosity profile of NGC 2070 (including R136) give a core radius of 
$\approx 0.4$ pc (Moffat \& Sweggewiss 1983).  Kennicutt \& Chu (1988), 
however, fit an isothermal profile to the stellar (rather than luminosity) 
distribution of NGC 2070 with a core radius of 4.5 pc when the contribution of 
R136 is ignored. Kennicutt \& Chu  question the significance of the 
similarity of NGC 2070 to a King model and more evolved globular clusters, 
doubting that such a young object could have thermalised.  

\section{$N$-body simulations}

The $N$-body code used in this paper is based upon Aarseth's (1996) nbody2 
code with the addition of stellar evolution and a variable external 
potential to model the effects of residual gas loss.  This code has been 
also been described in more detail in paper I.

A variety of initial conditions were tested to see which combinations 
produce clusters similar to those observed in the LMC.

In the simulations the stars and gas were originally distributed in a 
Plummer model.  Plummer models were chosen for their simple form which 
allows the forces due to the gas distribution to be easily calculated.  The 
choice of a Plummer model was not entirely arbritary, however.  Although 
NGC 2070 is inhomogeneous with little spherical symmetry (see section 
2.3) it does appear that, as a first approximation, it can be 
considered as a simple (isothermal) distribution, such as a Plummer 
model.  In the absence of a good model to explain globular cluster formation 
in detail, and with the limited resolution in such an $N$-body code as 
this, this approximation is used.  

The potential of a Plummer model is given by 

\begin{equation}
\phi (r) = - \frac{GM}{(r^2 + R_{\rm S}^2)}
\end{equation}

\noindent where $G$ is the Gravitational constant, $M$ is the total mass of 
the stars or residual gas and $R_{\rm S}$ is the length scale of the potential. 

The initial distribution function of the stars in a Plummer model is formed 
using the procedure detailed by Aarseth, H\'{e}non \& Wielen (1974) for  
random initial conditions.  The particles are distributed randomly in 
phase space in such a way as to produce a Plummer model with an isotropic 
velocity distribution.  The length and velocity scales can then be scaled so 
as to produce the desired scale length.  The velocities of the particles can 
be scaled to change the kinetic energy, $T$, of the system to produce 
the required virial ratio, $Q$, with the potential energy, $\Omega$, where 
$Q = T/\mid \Omega \mid = 0.5$ corresponds to a system in virial equilibrium.

\subsection{Tidal Field}

No tidal field was imposed upon the simulations in this paper.  The tidal 
field of the LMC is very weak and its extent and nature are poorly 
known (van den Bergh 1991).  As the clusters under investigation are all 
very young it is unlikely that they will have undergone any significant tidal stripping.  A possibly significant number of stars in the cluster may have 
overflowed the tidal boundary in this time which will be stripped over time 
and whose stripping may unbind the cluster leading to its eventual 
disruption.  In the timescales considered in this paper, however, this 
effect is assumed to be negligible  in the central $\approx 200$ pc.

\subsection{Initial Mass Function and Stellar evolution}

The initial mass function (IMF) is taken to be a power law of the form 
 
\begin{equation}
N(M) \propto M^{-\alpha}  
\end{equation}

\noindent where this corresponds to the Salpeter (1955) IMF for the solar 
neighbourhood when $\alpha=2.35$.  In these simulations $\alpha$ is 
taken to be 1.50 or 2.35.  These values are consistent with   
observations of the mass function in these clusters (section 2.3) which 
is presumably a good indicator of the IMF as these clusters have had 
little time to evolve in any way which may effect the mass function (see 
section 4.3).  The lower slope of 1.50 was chosen as the disruptive effects 
of mass loss from clusters with IMF slopes lower than this will 
certainly disrupt the cluster on small timescales (Chernoff \& Weinberg 
1990). 

Simulations were run with equal-mass particles to reduce the effects of 
relaxation which become apparent in multi-mass simulations (Giersz \& 
Heggie 1996).  This is different to the method used in paper I which 
used multi-mass models (cf. Fukushige \& Heggie 1995).  The quantitative 
results of small $N$ simulations using either method are similar but 
the exact form of profiles differs between the two methods.

The mass loss from stars due to stellar evolution is included in the code.  
Each particle represents a group of stars containing the full range 
of masses. Every Myr the masses of all the particles are changed by an amount 
representing the mass loss from stellar evolution in that Myr as some stars 
evolve to their appropriate end-state.  High mass stars ($M>8\msun$) become 
type II supernovae and evolve into neutron stars of 1.4$\msun$.  Intermediate 
mass stars ($4\msun < M < 8\msun$) have two possible end states.  They may 
become type I$^{1}\!/_{2}$ supernovae which totally destroy themselves, 
leaving no remnant (Iben and Renzini 1983), or they may evolve into white 
dwarfs of $\approx 1.4 \msun$.  The final stages of the evolution of 
these intermediate mass stars is poorly understood and so either possibility 
may be used in the code.  The difference in remnant mass does not appear to 
be especially important,  although it may affect the evolution of 
clusters with a low IMF slope which have significant numbers of 
intermediate mass stars.  Low mass stars, $2\msun < M < 4\msun$ evolve 
into white dwarfs of mass $0.58 + 0.22(m-1)$, where $m$ is the initial mass 
of the star, all in solar masses (Iben \& Renzini 1983).  Runs cover, at the 
very most, a few hundred Myr, normally 100 Myr, and so the evolution of 
stars less than $\approx 5\msun$ is not relevant.

The end-time of stellar evolution is given by fitting a line to 
the stellar evolutionary calculations of Maeder \& Meynet (1988)

\begin{equation}
{\rm Log_{10}} \left( \frac{M}{M_{\odot}} \right) = 1.524 - 0.370{\rm Log_{10}} 
\left( \frac{T}{\rm Myr} \right)
\end{equation}

\noindent These calculations are for solar metallicity stars.  The 
young LMC globular clusters are, at most, 20 times less abundant in metals 
than solar ([Fe/H]$> -0.7$) and hence these models are expected to be 
reasonably good approximations to the evolution of such stars.   

The numbers of stars reaching their end-states at any time is calculated from 
the slope of the IMF, $\alpha$.  If $N_{\rm M_{1} \rightarrow M_{2}}$ is the number 
of stars in the mass range $M_1$ to $M_2$ in a cluster with initial 
stellar mass $M_{\rm cl}$ then    

\begin{equation}
N_{\rm M_{1} \rightarrow M_{2}} = M_{\rm cl} \frac{(\alpha - 2)}{(\alpha - 1)} \frac{(M_{1}^{-(\alpha - 1)} - M_{2}^{-(\alpha - 1)})}{(M_{\rm low}^{-(\alpha - 2)} - M_{\rm up}^{-(\alpha - 2)})}
\end{equation}

\noindent (cf. equation 7 in paper I).  The upper limit of the IMF, 
$M_{\rm up}$ is 
taken to be $15\msun$.  The low mass end of the IMF, $M_{\rm low}$ is set 
at $0.15\msun$ corresponding to the observed turn over  
in the mass function in observations of the Galactic globular cluster NGC 
6397 (Paresce, De Marchi \& Romaniello 1995).

The mass lost by stellar evolution is ignored.  Prior to the gas expulsion 
the mass lost is assumed negligible compared to the mass of residual gas 
(although it is often the driving force behind the residual gas expulsion).  
After the residual gas has been expelled, any mass loss from stellar 
evolution is assumed to be energetic enough to leave the cluster and not 
massive enough at any time to affect the dynamics of the stars in the 
cluster.  This assumption in young clusters with low slopes to their mass 
functions may not be realistic, but is used for simplicity.

\subsection{Star Formation Efficiency}

The SFE of a globular cluster is one of the most important initial 
conditions included within this study.  The SFE is defined simply as the 
fraction of the initial mass of gas which is turned into stars.  In the 
Galactic disc star forming regions are usually found to have SFEs of order a 
few percent (Larson 1986).  A simple application of the virial theorem 
leads to the conclusion that for bound clusters to form and survive the 
expulsion of residual gas from star formation then an SFE of {\em at least} 
50\% is required.  Lada \et (1984) showed that open clusters can retain a 
bound core of stars with SFEs as low as 30\%.  Paper I suggested that, if 
the initial concentration of a globular cluster is high enough, then it 
could survive the expulsion of residual gas with SFEs possibly as low as 
25\%.  This is possible if the cluster stars are able to settle 
into a new (larger) dynamical equilibrium which is well inside the tidal 
limit of the cluster.  Clusters were given a variety of SFEs from 10\% to 
80\%.  As shown in paper I, the effect of residual gas expulsion from a 
cluster is highly dependent upon the SFE. 

\subsection{Residual gas expulsion}

The expulsion of the residual gas in the cluster not used in star 
formation is modelled by a variable external potential acting upon the 
particles in the cluster (cf. Lada \et 1984).  The 
residual gas initially has a Plummer distribution, chosen due to the 
simple analytic form of its potential.  The Plummer model is given the 
same scale length as that of the stars, but often a different mass (depending 
upon the SFE of the cluster in question).  The effect of stars upon the 
residual gas is assumed to be restricted to the expulsion of the gas as the 
gas is being heated by the massive stars and is present for a relatively 
short length of time.

In this paper two simplified mechanisms are used to expel residual gas 
from a globular cluster.  The first mechanism is based upon the gradual 
depletion of the gas in a cluster by the action of the UV fluxes and 
stellar winds from massive stars (based upon the hydrodynamic simulations of 
Tenorio-Tagle \et 1986).  It may also include some gradual expulsion by 
supernovae which would be expected if star formation is not virtually 
instantaneous.  This expulsion is simulated by the slow, constant reduction 
(on a timescale of a few Myr) of the mass of gas in the cluster to zero 
starting a few Myr after the end of star formation.  This type of 
residual gas expulsion was found by paper I to be the least 
disruptive to a cluster.  This is due to the gradual and global nature of the 
change in potential which is slow compared to a normal crossing time 
allowing the stars to adjust gradually.

The second mechanism of residual gas expulsion is by the additive action of 
large numbers of supernovae near the centre of the cluster to form a 
'supershell' (of the type observed in OB associations in the Galaxy) 
which sweeps the cluster clear of gas (Brown, Burkert \& Truran 1995).  
This supershell is modelled by assigning the supershell a radius $r_{\rm 
shell}$ at some time $t$ depending upon the rate of supernova events $\dot{N}$ 
assumed to be constant during the time supernovae occur, the energy of 
each supernova $E$ taken to be $10^{51}$ ergs and the external gas 
pressure $P_{\rm ext}$ on the shell.  This is given by (equation (9) in 
Brown \et 1995)

\begin{equation}
r_{\rm shell} = \left( \frac{3}{10\pi} \frac{E\dot{N}t}{P_{\rm ext}} \right) 
^{^{1}\!/_{3}}
\end{equation}
 
If a star is interior to $r_{\rm shell}$ then it feels no force due to the 
gas while exterior to $r_{\rm shell}$ it feels the same force as if 
the supershell were not there (from Newton's first theorem).  This mechanism 
of residual gas expulsion was found by paper I to be the most disruptive to 
the cluster as the shell sweeps out the inner few pc (where most of the stars 
are present) on a timescale far less than a crossing time.  As the gas 
expulsion from the inner regions is so fast the assumption that all the 
supernovae occur at the centre of the shell is probably fair.  On such a 
timescale the stars are unable to adjust to the change in potential gradually 
which results in a more dramatic change.

This mechanism is of particular interest as studies of the dynamics of the 
gas in the 30 Doradus HII region around the cluster NGC 2070 (which 
contains the group of massive stars R136) indicate that this region will 
eventually become a supershell (Chu \& Kennicutt 1994).  This study 
also underlines, however, the extremely complex nature of the gas around 
a large, young star cluster which a simulation such as this 
could not attempt to model accurately.

In paper I it was mentioned that for low SFEs not enough high mass stars 
might be present to expel the residual gas.  This consideration is not 
of importance in this paper due to the low values of $\alpha$ of the IMFs.  
Such values of $\alpha$ should always produce enough high mass stars to 
expel the residual gas for the SFEs of interest.

\subsection{Core radii}

The core radii of clusters is an important observational quantity.  In order 
to compare simulations and observations the core radii are calculated in 
two ways.  The first is based on the method described in Casertano \& 
Hut (1985).  This method is designed to produce core radii from $N$-body 
simulations which can be compared to observational core radii.  The core 
radius, $r_{\rm c}$, is calculated by 

\begin{equation}
r_{\rm c} = \left( \frac{\sum_{i=1}^{n} \mid {\bf r}_i - {\bf r}_d \mid ^2 \rho_i^2}{\sum \rho_i^2} \right) ^{1/2}
\end{equation}

\noindent where $\rho_i = 15/(4\pi r_6^3)$ is the density estimator of 
particle $i$ with respect to the sixth  nearest particle $r_6$ and ${\bf r}_d$ 
is the position of the density centre.  In nbody2 the original procedure has 
been modified to sum over a central sample of $n \approx N/2$ particles. 

This core radius is compared to the core radius obtained from the 
fitting of the mass profile derived from equation 1.  Fittings may be 
made with one or two free parameters: $\gamma$ and $a$ (related to 
$r_{\rm c}$); or, assuming $r_{\rm c}$ to be given by equation 7, only 
$\gamma$ as a free parameter.  These methods usually produce very similar 
results.  Some discrepancy can occur due to the sometimes rapid variations 
in $r_{\rm c}$ from equation 7 caused by the relatively small number 
of particles (1000) used in most simulations. 

\subsection{Computational parameters}

Heggie \& Mathieu's (1986) standard $N$-body units were used where $M_0=G=1$ 
and $E_0=-1/4$, where $M_0$ is the initial mass of the particles (ie. 
the initial stellar mass of the cluster) and $E_0$ is the initial energy of the 
particles.  Conversion to units of time (in Myrs) for the treatment of 
stellar evolution was made using the relationship

\begin{equation}
T_{\rm c} = M^{5/2}_0 / (2 \, \vert E_0 \vert)^{3/2} = 2\sqrt 2 U_t  
\end{equation}

\noindent where $U_t$ is the unit of time within the code.

The softening length in all simulations was set to be 1/50.  Equal-mass 
simulations run with softenings between 1/20 and 1/100 show similar 
results.  This level is softening is less than that used in multi-mass 
simulations where the effects of relaxation are stronger.  

\begin{figure}
\centerline{\psfig{figure=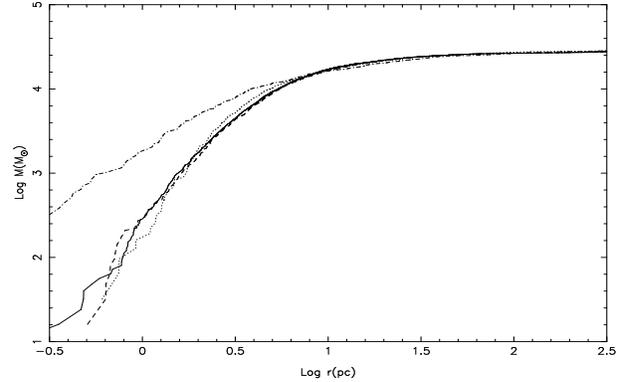,height=5.0cm,width=8.0cm,angle=270}}
\caption{Mass profiles after 100 Myr for four identical clusters with 
$R_{\rm S}$ = 2.4 pc, $\alpha = 1.50$ and an SFE of 40\% with particle 
numbers $N$ = 500 (dot-dash line), 1000 (dotted line), 2000 (dashed line) 
and 4000 (solid line).}
\label{figure:pnumber}
\end{figure}

The effect of varying the number of particles, $N$, used in the simulations 
is illustrated in fig.~\ref{figure:pnumber}.  As can be seen, the profiles 
for $N \gte 1000$ are very similar, profile fits to all 3 curves gives 
a fit of $\gamma = 2.9 \pm 0.1$ and $r_{\rm c} = 2.6 \pm 0.2$ (errors 
come from combining the 3 different fits, they are not the errors inherent in 
each individual fit).  Note also that the mass interior to any radius is 
virtually the same for $N \gte 1000$ runs when $r \gte 1$ pc.

Most simulations were made with $N=1000$ due to the large number of 
simulations and limits on available CPU time.  To test the statistical 
validity of results (Heggie 1995) most simulations were made 3 or 4 times.  
The initial conditions were kept constant but the random numbers used 
to generate the distribution function of the stars were given a different 
`seed'.  A few simulations were repeated more to further reduce statistical 
noise. 

\section{Results}

The results of the $N$-body simulations and their comparisons to 
observation are presented in this section.  In section 4.1 the general 
results on the shape and evolution of the profiles of varying different 
initial conditions are given.  A  model for the initial conditions 
of these young globular clusters is suggested.  In section 4.2 the 
implications of these results for the survivability of these clusters both 
in the LMC and in the Galactic environment are discussed.  Finally in 
section 4.3 the effect of the IMF slope upon cluster evolution and 
survivability is considered.

\subsection{The initial conditions of young LMC globular clusters}   

The aim of the simulations was to find the initial conditions that most 
effect the shape of the mass distribution profile after gas expulsion and 
produce the range of cluster parameters presently observed.  The 
initial conditions that were varied to try and recreate these profiles were 
the mass, SFE, IMF slope, initial distribution function of the stars and 
gas expulsion mechanisms. The affects upon the mass distribution profile 
of altering each of these initial conditions are discussed individually 
below followed by a discussion of the set of initial conditions that 
would result in the variety of young globular clusters in this sample. 

When fitting profiles to the mass distributions of the simulations care was 
taken to make the fitting similar to that of EFF.  The fits were made to 
a log-log profile of mass-radius and fitted primarily in the inner few 
pc of the simulations.  The profile within $\approx 1$pc was often not used 
to fit as it is highly sensitive to random variations in particle number due 
to the low numbers of particles used.  Fits were made out to 
$\approx 200$ pc, a distance similar to that out to which the observational 
fits were made. 

\begin{figure}
\centerline{\psfig{figure=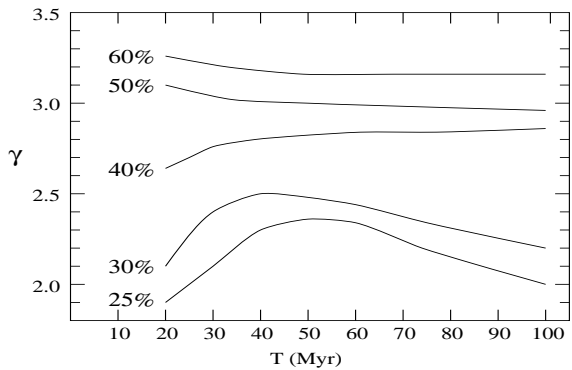,height=5.0cm,width=8.0cm,angle=270}}
\caption{The evolution of $\gamma$ over the first 100 Myr for different star 
formation efficiencies for clusters which expel their residual gas 
via a supershell.  These clusters had an initial mass of $5 \times 10^4 \msun$.}
\label{figure:gt}
\end{figure}

The SFE is found to have the most influence upon the mass profile of 
the cluster after gas expulsion.  The lower the SFE of a cluster, the 
smoother its profile will be (ie. the value of $\gamma$ required to fit 
the profiles of low SFE clusters is lower).  The effect of the SFE 
upon $\gamma$ during the first 100 Myr is shown in fig.~\ref{figure:gt}. 
It is found not to vary significantly with the chosen IMF or Plummer model 
scale length in any simulation.  The lines on fig.~\ref{figure:gt} are 
guides only: an average over many simulations.  In any particular 
simulation the fitted value of $\gamma$ may vary by 0.3 from the plotted 
values.  The general trend is the same, however, in the majority of runs.

\begin{figure}
\centerline{\psfig{figure=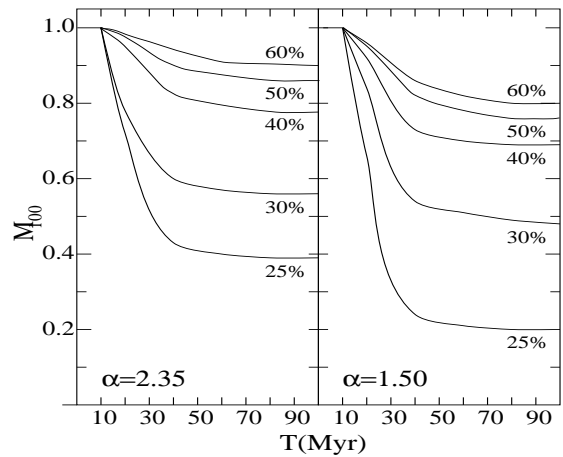,height=6.0cm,width=8.0cm,angle=270}}
\caption{The change in the fraction of the initial stellar mass of a 
cluster remaining within 100 pc of the centre of mass with time for 
different SFEs from 25\% to 60\% (as marked).  These 
clusters were initially $5 \times 10^4 \msun$ with $R_{\rm S} = 2.4$pc 
and $\alpha = 2.35$ or $\alpha = 1.50$.  In all cases the residual gas 
was expelled by a supershell mechanism after 9 Myr.}
\label{figure:mt}
\end{figure}

\begin{figure}
\centerline{\psfig{figure=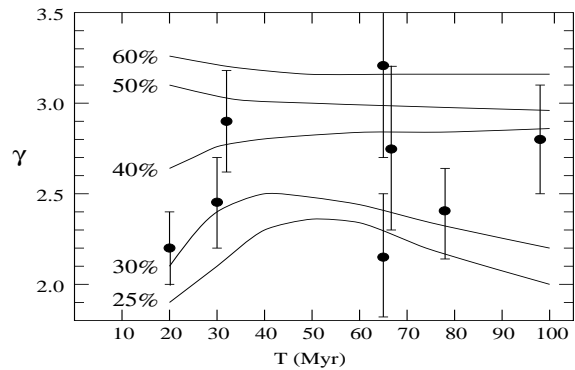,height=5.0cm,width=8.0cm,angle=270}}
\caption{The T-$\gamma$ evolution for different SFE with observations (from 
tables 1 and 2) overlaid including the error bars for $\gamma$.  Note that 
the  x-position of NGC 2156 has been shifted slightly to the right to allow 
its error bars and those of NGC 2159 and 2172 to be seen more clearly.}
\label{figure:gtobs}
\end{figure}

Figure~\ref{figure:mt} shows that clusters with a low SFE lose a 
high proportion of their stars rapidly once gas expulsion is complete.  
Many stars are unbound by the loss of the residual gas and escape from 
the cluster.  This escape is not instantaneous, however, and during the 
10 to 20 Myr required for these stars to escape the cluster profile 
appears to have a very low $\gamma$ (see fig.~\ref{figure:gt}), a feature 
which appears to occur in the observations (see fig.~\ref{figure:gtobs}).   
In clusters with high SFE the effect of the residual gas expulsion is not 
felt so strongly by the cluster and the immediate escape of stars after 
gas expulsion is not so dramatic. It should be noted that low escape rates 
in the first 100 Myr do not necesserally produce a bound cluster that 
will survive.  The further effects of stellar evolutionary mass loss and 
gradual evaporation of stars could well disrupt these clusters (Chernoff \& 
Weinberg 1990).

Altering the total initial stellar mass of a cluster is found to have a very 
small effect upon the resulting shape of the profile.  Lower initial mass 
clusters appear to have a smoother mass distribution than those of a 
higher mass.  However, this effect is very small and is swamped by the effect 
of changing the SFE.  This would explain the lack of any correlation 
between cluster mass and $\gamma$ in the sample.

The cluster mass will affect the survivability of a globular cluster as 
tidal overflow after residual gas expulsion is more disruptive in 
low mass clusters (paper I).  It is this tidal overflow that is assumed 
to produce the extensive unbound stellar halos observed in these young LMC 
globular clusters (EFF).

The mechanism of gas expulsion does also have an effect upon the shapes 
of the mass profiles.  This effect is, again, 
small compared to the changes caused by different SFEs.  The more disruptive 
supershell expulsion, unsurprisingly perhaps, forms clusters with lower 
$\gamma$ fits (smoother 
mass profiles) than the more sedate gradual expulsion mechanism.  In addition, 
supershell expulsion causes the unbinding of a higher proportion of a 
cluster's stars and the formation of a very large halo of stars.

The observation that these young globular clusters do have large, 
unbound halos (EFF, van den Bergh 1991) combined with the evidence that 
NGC 2070 will eventually form a supershell (Chu \& Kennicutt 1994), would 
appear to suggest that a supershell mechanism for the expulsion of the 
residual gas may be the better approximation.

The initial virial ratio (initial energy distribution) of the stars in 
the cluster is found to have a large affect upon the form and evolution 
of profiles during the first 100 Myr of cluster evolution.  The 
value of $Q$ can greatly effect the levels of stellar escape after gas 
expulsion as well as the core radius of the cluster.  The virial ratio 
of the cluster quickly settles down to its equilibrium value of $Q=0.5$ for 
unvirialised clusters (see also paper I).  In attaining this equilibrium, 
however, the cluster must change its distribution (expanding if 
$Q>0.5$, and contracting if $Q<0.5$).

\begin{figure}
\centerline{\psfig{figure=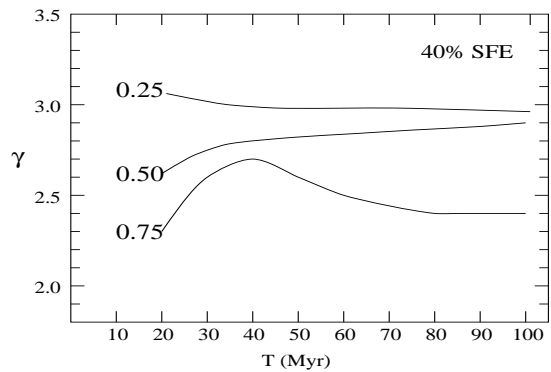,height=5.0cm,width=8.0cm,angle=270}}
\caption{The effect of the initial virial ratio $Q$ upon the evolution of 
the profile parameter $\gamma$ for $Q$=0.25, 0.50 and 0.75.  All 3 clusters 
have an SFE of 40\% initially with $R_{\rm S}=2.4$, $M_{\rm tot} = 
5 \times 10^4 \msun$ and $\alpha = 2.35$.}
\label{figure:gtvir}
\end{figure}

Figure~\ref{figure:gtvir} shows the effect upon the evolution of $\gamma$ 
of varying the initial virial ratio.  For $Q < 0.5$ the form 
and timescale of the escape is not altered significantly so that the  
evolution of $\gamma$ with SFE remains approximately the same.  When $Q>0.5$ 
the cluster's evolution can be significantly changed.  The change of $\gamma$ 
observed in fig.~\ref{figure:gtvir} when $Q=0.75$ corresponds to a far 
less bound cluster which loses a larger number of its stars immediately 
after gas expulsion as the velocity dispersion of the cluster is higher 
(producing a similar evolutionary profile to that of lower SFEs 
in fig.~\ref{figure:gt}).  After 100 Myr the $Q=0.25$ and $Q=0.50$ clusters 
both lose $\approx 15\%$ of their particles beyond 100 pc from the 
cluster core, in the $Q=0.75$ case, however, nearly 40\% of the particles 
have passed the 100 pc radius in the same time.  

\begin{figure}
\centerline{\psfig{figure=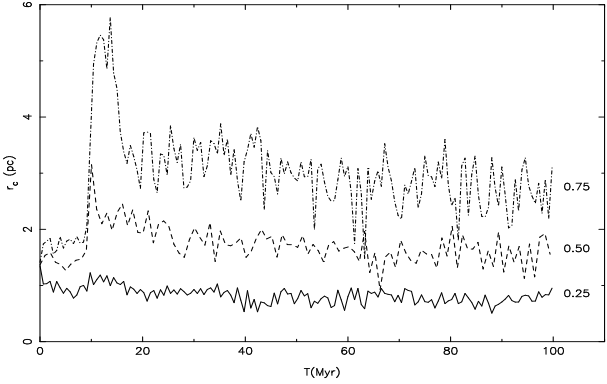,height=5.0cm,width=8.0cm,angle=270}}
\caption{The effect of the initial virial ratio $Q$ upon the evolution of the 
core radius $r_{\rm c}$ (calculated from equation 7.) for $Q$=0.25, 0.50 
and 0.75.  The initial 
conditions of the clusters were the same as those in fig.~\ref{figure:gtvir}.}
\label{figure:rcvir}
\end{figure}

The strong dependence of core radius on the initial virial ratio is shown 
in fig.~\ref{figure:rcvir}.  The profiles are unsmoothed and show the 
variations in $r_{\rm c}$ caused by the `noise' introduced by the low 
particle number (1000).  The expansion to equilibrium and the more 
disruptive effect of gas expulsion in the $Q=0.75$ cluster is shown in 
the far higher core radius of that cluster compared to the other clusters.  
In addition this cluster shows a huge expansion of $r_{\rm c}$ caused by 
the act of residual gas expulsion around 9 Myr.  

The initial concentration of clusters in the simulations is not an entirely 
free parameter.  The core radii of the simulated clusters have to 
correspond with the observed range of core radii $1 < r_{\rm c} < 3.5$ 
after residual gas expulsion.  This limitation is found to place 
surprisingly strict bounds upon the allowed range of initial stellar 
distributions (in these cases, the scale length $R_{\rm S}$ of the 
Plummer models).  The core radius of a cluster may increase after gas 
expulsion by a factor of $\approx 2$ to $4$ for clusters with low SFEs 
before contracting (see fig.~\ref{figure:rcvir}).  
Residual gas expulsion from clusters with high SFEs ($>40\%$) does not 
appear to have a significant effect upon the core radii of the simulations. 
However, soon after gas expulsion is completed, the central regions 
of clusters will contract unless the IMF slope is low enough (this is 
discussed in detail in section 4.3).  In order for the simulations core 
radii to remain in the observed range over the first few hundred Myr of 
evolution then the scale length of the Plummer models (related to core 
radius by $r_{\rm c}=0.644R_{\rm S}$) is found to have to lie in the range 
1 to 3 pc, depending to some extent upon the IMF slope.  

\begin{figure}
\centerline{\psfig{figure=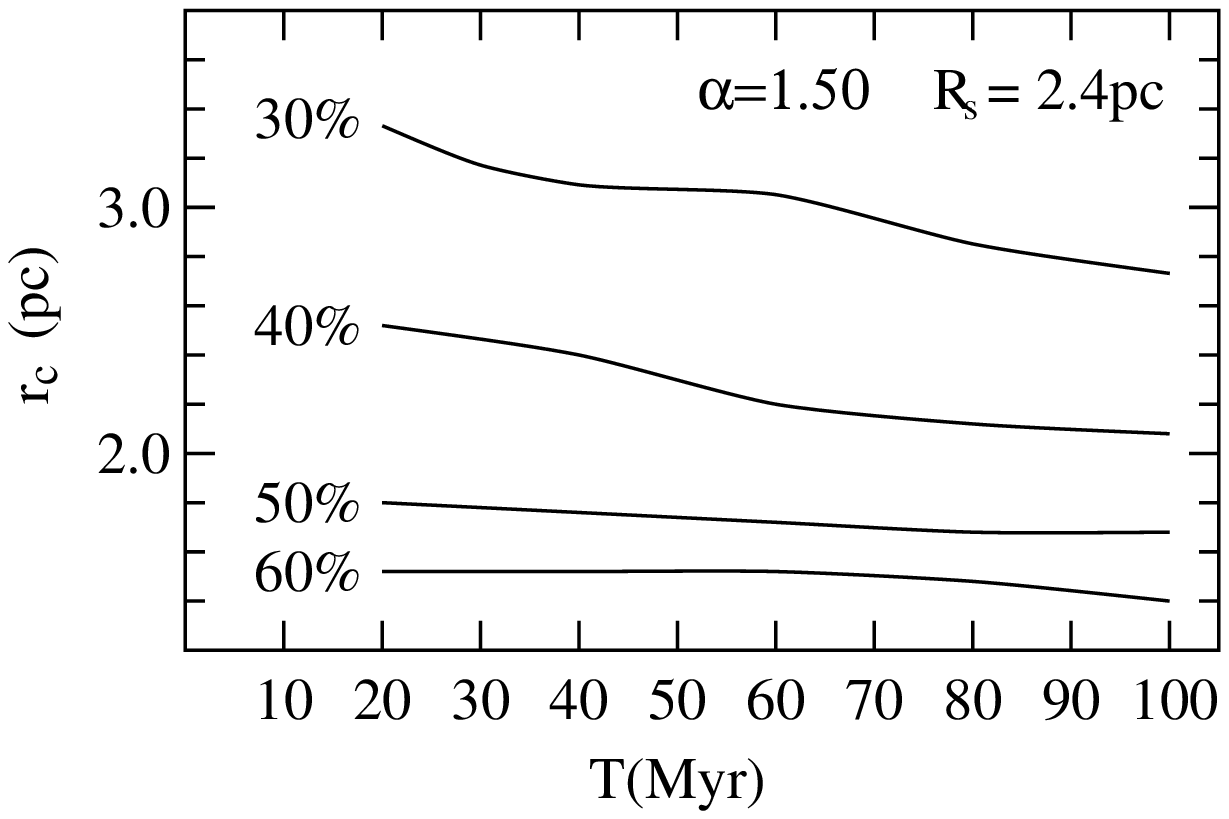,height=5.0cm,width=8.0cm,angle=270}}
\centerline{\psfig{figure=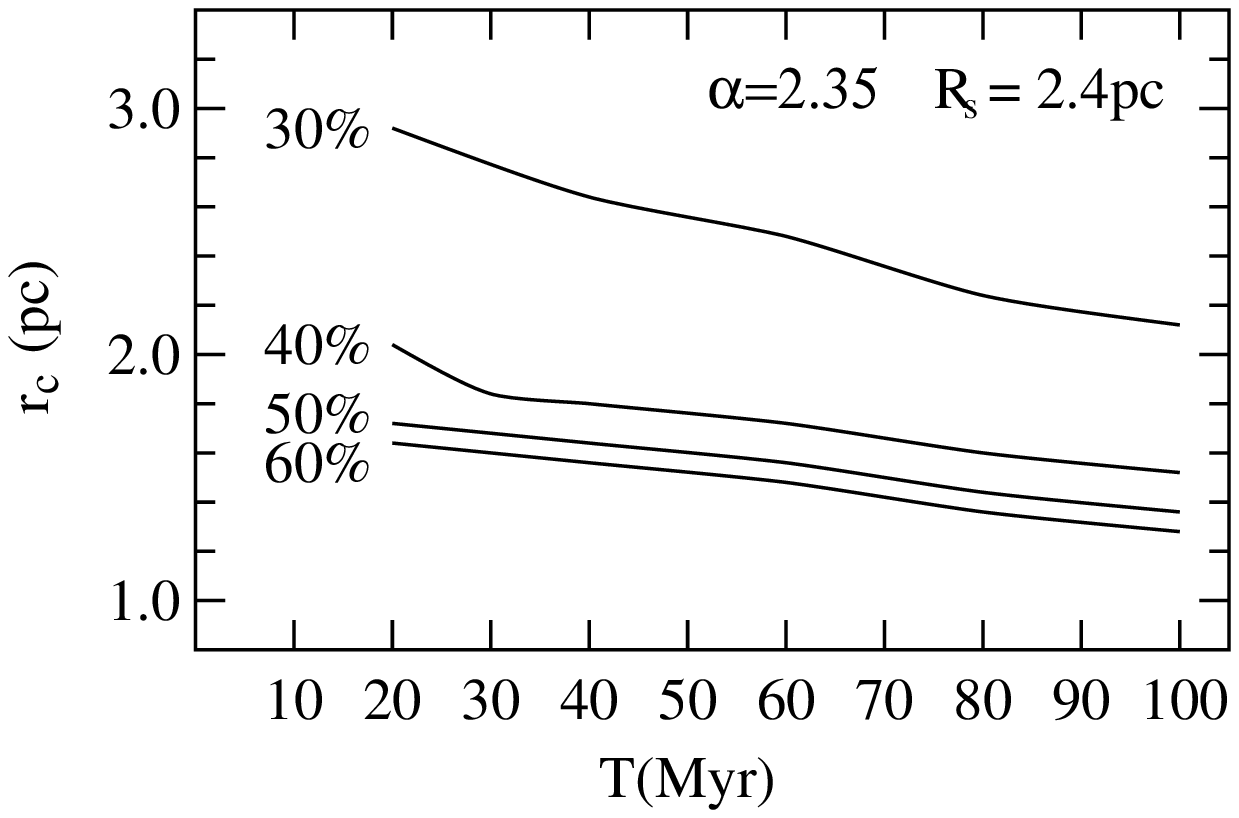,height=5.0cm,width=8.0cm,angle=270}}
\caption{The evolution of core radius with different SFEs for $R_{\rm S} = 
2.4$pc clusters with IMF slopes of $\alpha = 2.35$ (top) and 1.50 
(bottom).  The core radii are smoothed averages. }
\label{figure:rct4}
\end{figure}

\begin{figure}
\centerline{\psfig{figure=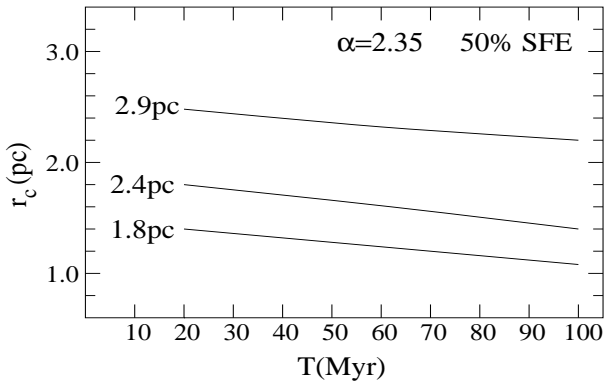,height=5.0cm,width=8.0cm,angle=270}}
\caption{The change in the evolution of core radius for a 50\% SFE, 
$\alpha = 2.35$ cluster with different initial Plummer model length scales. 
The core radii are smoothed averages.}
\label{figure:rctcomp}
\end{figure}

Figure~\ref{figure:rct4} shows how the evolution of the core radius is 
strongly dependent upon the SFE of the simulation.  Note that the 
core radii in figs.~\ref{figure:rct4} and~\ref{figure:rctcomp} are smoothed 
averages over several simulations.  In any one simulation $r_{\rm c}$ 
fluctuates by $\approx \pm 0.5$ pc and the average value of $r_{\rm c}$ 
may differ between simulations by about the same amount.  These lines should therefore only be taken as guides as to the behaviour of $r_{\rm c}$. 
The decline in $r_{\rm c}$ over the 
first 100 Myr of the clusters evolution is due to the resettling of the 
bound cluster core after gas expulsion.  Such reductions are also observed in 
many $\alpha = 1.50$ simulations which begin to re-expand from the effects of 
stellar evolutionary mass loss after a few hundred Myr.

The fraction of particles within $R_{\rm c}$ is also found to decline 
after residual gas expulsion.  The level of this decline is also 
stronger for lower SFEs.  Initially $\approx 10\%$ of particles are 
within $r_{\rm c}$ this declines after 100 Myr to $\approx 7\%$ with an 
SFE of 50\% and $approx 4\%$ if the SFE is 30\%.  The reduction is 
sligtly ($\approx 1\%$) greater for $\alpha = 1.50$ as opposed to $\alpha = 
2.35$.

The dependence of $r_{\rm c}$ upon the scale length of the original Plummer 
model is obvious in fig.~\ref{figure:rctcomp}.  The sample SFE of 50\% is 
chosen to illustrate the general behaviour of $r_{\rm c}$ with $R_{\rm S}$.  
Very low SFEs ($<30\%$) rise more dramatically with increasing $R_{\rm S}$, 
limiting the range of allowable Plummer models more.

It is found that the profiles of the youngest clusters are often not smooth 
and easily fitted by the EFF profiles.  This appears to be due to the loss of 
stars immediately following gas expulsion, especially in 
low SFE clusters.  The bumps, steps, and shoulders noted 
by Elson (1991) in the luminosity profiles are present in the mass profiles 
of simulated clusters.  These departures from the EFF profiles are at their 
most extreme for a few tens of Myr after the completion of residual gas 
expulsion.  They are gradually smoothed out by dynamical processes and the 
loss of high velocity particles from the cluster.  Some of the observed 
structure may be due to the poor resolution of the $N$-body simulation,  
although such structures are observed in actual clusters.

The required 
length scales for young clusters ($<50$ Myr) are generally independent of 
the IMF slope.  In these young clusters $\gamma$ and $r_{\rm c}$ are 
determined more by the results of the residual gas expulsion than any 
internal dynamical process such as stellar evolutionary mass loss.  In older 
clusters ($>50$ Myr), however, the additional evolutionary effects of 
a low IMF slope are being felt by the cluster.  This results in higher 
$r_{\rm c}$ for a given $R_{\rm S}$ and SFE. 

Figure~\ref{figure:gtobs} shows the evolution of $\gamma$ for a variety of 
SFEs with the observations overlaid.  The quoted errors for $\gamma$ from 
EFF have been included. Table 3 shows the implied initial 
conditions for the observed young LMC globular clusters assuming they 
are initially virialised ($Q=0.50$) combining fig.~\ref{figure:gtobs} and 
the results for the evolution of $r_{\rm c}$ with SFE and $R_{\rm S}$. 
As can be seen from fig.~\ref{figure:gtobs} the 
quoted SFEs in table 3 contain an error of $\approx \pm 10\%$ due to the 
errors in $\gamma$.  Large error bars also exist for the 
ages of the clusters, however these errors would not introduce such a 
variation into the possible SFEs.  Table 
3 combines the results from fig.~\ref{figure:gtobs} with values for 
$r_{\rm c}$ for different Plummer model $R_{\rm S}$ to estimate the 
initial conditions that would result in the observed clusters with IMF 
slopes of $\alpha = 2.35$ and $1.50$.  

The values of $R_{\rm S}$ obtained for either IMF slope are the same for young clusters whose profile is determined by the effect of the residual gas 
expulsion, older clusters tend to have a lower $R_{\rm S}$ for $\alpha = 
1.50$.  The quoted $R_{\rm S}$ should be taken as a guide only and 
contain an error of at least $\pm 0.5$ pc.  It should be noted that they 
are for virialised clusters which have have their residual gas 
expelled by a supershell.  Non-virial equilibrium will alter this value (decreasing for $Q>0.5$ and increasing for $Q<0.5$) and less disruptive 
gas expulsion mechanisms will increase this value.  The relative values will 
stay approximately the same, however.

Combining the above results as to the effects of various initial 
conditions upon the mass distributions of young clusters it is 
suggested that the  population of young LMC globular clusters could have 
been formed from a fairly uniform population of proto-cluster clouds.  The 
initial star formation forms a fairly relaxed Plummer-like distribution 
from which residual gas is expelled by the actions of supernovae.  The 
differences in cluster profiles could be explained as differences in the mass 
and SFE of the proto-cluster.  

\begin{center}
\begin{table}
\begin{center}
\begin{tabular}{|c|c|c|c|} \hline
NGC & SFE & $R_{\rm S}$ (pc)  & $R_{\rm S}$ (pc)  \\  
    &     & $\alpha = 2.35$   & $\alpha = 1.50$ \\ \hline
1818 & 30\% & 1.8 &  1.8 \\   
2004 & 30\% & 1.2 &  1.2 \\ 
2156 & 40\% & 2.4 &  2.4 \\ 
2157 & 45\% & 2.9 &  1.8 \\ 
2159 & $>$25\% & 1.8 & 1.2 \\ 
2164 & 40\% & 2.4 &  1.8 \\ 
2172 & 60\% & 2.9 &  2.6 \\ 
2214 & 30\% & 2.4 &  1.8 \\ \hline
\end{tabular}
\end{center}
\caption{The star formation efficiencies and Plummer model scale lengths for 
IMF slopes $\alpha$=2.35 and 1.50 required to match the observations 
of the sample clusters younger than 100 Myr.  All these clusters were 
initially in virial equilibrium and the residual gas was expelled by a 
supershell beginning at 9 Myr.}
\end{table}
\end{center} 

This model provides an explanation for the differences observed in the 
quartet, in particular the coeval clusters NGC 2156, 2159 and 2172.  These 
clusters, despite their striking similarities in age and location are 
very different in their profiles.  They have the widest range of profile 
shape observed in the entire sample with $\gamma$ = 2.75, 2.15 and 3.20 
respectively.  If the SFE is the major factor in 
the determination of profile shape in these clusters then this discrepancy 
could be accounted for by varying the SFE between 25\% and 60\% in these 
clusters.  The derived scale length for these clusters is $\approx 2.4 \pm 
0.5$ pc.  A constant scale length of 2.4 pc is possible within the errors 
quoted for $\gamma$.  Given this a common origin for these clusters would 
appear plausible.

EFF also determined profiles for 2 other young LMC globular clusters 
NGC 1831 and NGC 1866.  Both of these clusters have not been included in 
the main analysis as they are older than 100 Myr, ie. 400 Myr and 138 Myr 
respectively (Girardi \et 1995).   NGC 1866 would not be expected to have had 
a strong tidal influence from the LMC, but NGC 1831 may have had some 
significant stripping of its tidal overflow modifying its profile. 
NGC 1866 fits well into the model of initial conditions, its profile 
being explicable with a scale length of $\approx 3$ pc and a 40\% SFE.  
NGC 1831 is observed to have $\gamma = 3.35 \pm 0.56$ which would seem 
to imply a high SFE, however, the extent to which its profile would have 
been modified by the tidal field is difficult to quantify.  The core radius 
of NGC 1831 is also large (3.3 pc) but may have been substantially altered 
if the cluster had a low slope to its IMF.

\subsection{The survivability of the young LMC globular clusters}

If the conclusions drawn in the previous section are a fair representation 
of the initial conditions that have produced the young LMC globular 
clusters then it should be possible to say something about the survivability 
of those clusters.  Paper I presented results as to what initial conditions 
are necessary to produce a globular cluster that will survive for a 
Hubble time in the Galaxy.  These conditions may be applied to the clusters 
in this sample to assess their survival chances.

Conditions within the LMC, especially with respect to the far weaker tidal 
field when compared to the Galaxy, will mean that the survival conditions of 
paper I are an underestimate of the survivability of the young LMC globular 
clusters.  However, it is of interest to 
examine whether these clusters would be able to survive within the Galaxy 
to test if they are, indeed, representative of younger versions of the 
old Galactic globular cluster population.

Interestingly, these clusters lie around the border-line of survival 
from paper I.  A cluster with an initial mass of $10^4 \msun$ 
with a Galactocentric distance of 5 to 8 kpc when $\alpha = 2.35$ is 
estimated to be able to survive if its initial Plummer model length scale 
$R_{\rm S} \lte 1.5$ pc for supershell gas expulsion for an SFE of 50\%.  
For an initial mass of $10^5 \msun$ this rises to $R_{\rm S} \lte 2.5$ pc. 
These values are somewhat lower than the scale lengths observed in the 
simulations which recreate the observed profiles.  However, for Galactocentric 
distances greater than $\approx 12$ kpc some of the clusters in this 
sample should be capable of surviving for a Hubble time if $\alpha \gte 2$.  
Clusters with IMF slopes of $\alpha = 1.50$ were not considered in paper I 
as Chernoff \& Weinberg (1990) showed that clusters with reasonable 
initial conditions could not survive for a Hubble time in the Galaxy with 
such IMF slopes.  

The range of scale lengths inferred for the initial Plummer models of these globular clusters implies an initial central 
density within the proto-cluster clouds in the range $10^3$ to $10^4 \msun$ 
pc$^{-3}$ - similar in range to that observed in giant molecular clouds 
in the Galaxy today (Harris \& Pudritz 1994).  This range is similar to 
that found as the border-line survival range for Galactic globular clusters 
in paper I.  This similarity with Galactic giant molecular clouds is also 
independent of the IMF slope.  

These values for the maximum scale length required for survival are also 
dependent upon the SFE of a cluster.  The higher the SFE, the less disruptive 
is the gas expulsion mechanism, and the higher the maximum scale length 
required for survival.  The clusters in the sample which require low SFEs 
to produce the observed profiles would not be expected to be able to 
survive within the Galaxy for a significant length of time.  Those clusters 
with high $\gamma$ and so, presumably, high SFEs may be able to survive for 
a Hubble time within the Galaxy.  

Considering the less disruptive nature of the environment around the LMC 
many of these sample clusters may be expected to survive for a significant 
length of time ($\gte 1$ Gyr).  NGCs 2156 and 2164 in particular 
would appear to be strong candidates for clusters that could survive for 
a significant time even within the Galactic environment if their IMF slopes 
are sufficiently high.

At the other extreme, NGCs 1818, 2159 and 2214 would almost certainly not 
be able to survive for any significant length of time in the Galaxy, and 
maybe not for more than a few hundred Myr even within the LMC.  Their 
low implied SFEs and high core radii are indicative of weakly bound clusters.

\subsection{The IMF slope}

The slope of the IMF significantly affects the structure and evolution 
of clusters.  The discussion of the affect of the IMF slope is included 
separately as it is the least well known of the structural parameters 
which will affect these clusters.  As noted in section 2.3 values for 
the IMF slope vary widely between clusters and the methods used to 
determine them, even within the same cluster.  None of these methods 
are entirely reliable and selection of the correct value of the IMF 
slope from the present literature appears impossible.  For these reasons 
the effect of the IMF slope is discussed for all apparently reasonable 
values.

A note of caution should be added here about the slope of the IMF.  The 
long-term survivability of globular clusters is highly dependent upon this 
quantity (Chernoff \& Shapiro 1987, Chernoff \& Weinberg 1990).  If the 
young LMC globular clusters do, indeed, have as low a slope to their IMFs 
as suggested by some studies (see section 2.2) then the possibility of 
their surviving for any significant time, even in the LMC, would appear 
minimal.  Chernoff \& Weinberg (1990) find that for IMF slopes lower 
than $\alpha = 1.5$ that a cluster cannot survive for a Hubble time with 
any reasonable initial conditions (and certainly not for the initial 
conditions implied for these clusters in this paper). 

It seems highly improbable that the slope of the mass function could 
have been altered by dynamical processes far from the slope of the IMF 
in such young clusters.  Multi-mass simulations including a tidal cut-off 
show a very minor change in the mass function after residual gas 
expulsion from the preferential escape of low mass particles ($\alpha = 
2.35$ drops to $\alpha \approx 2.25$).  It must be noted, however, that 
any mass effects would be considerably dampened by the significant softening 
used in the simulations.

\section{Discussion}

This paper presented the results of a comparison of observations of young 
($\lte 100$Myr) Large Magellanic Cloud (LMC) globular clusters with 
$N$-body simulations of 
globular clusters.  The $N$-body simulations were based on Aarseth's (1996) 
nbody2 code.  Stellar evolutionary 
mass loss and a variable external potential used to simulate the expulsion 
of residual gas (that gas not used to form stars in the initial burst of 
star formation in the globular cluster) were included in the code.

The aim of the paper was to explore the initial conditions which gave rise to 
these young globular clusters.  The initial conditions of 
particular interest were the initial spatial and energy distributions of the 
stars and the star formation efficiency (SFE) of the proto-cluster.  The 
initial mass function (IMF) slope was varied between 2.35 (the Salpeter 
value) and 1.50, covering the mid to upper end of observed values. 

The simulations were compared primarily with the observations of 
8 young LMC globular clusters made by Elson, Fall \& Freeman (1987, EFF).  
EFF found best fit luminosity profiles for these clusters and deprojected 
these to give density profiles.  These density profiles were integrated to 
obtain the mass distribution.  The observed and simulated mass distributions 
were compared to find which initial conditions gave a good representation 
of the observations.

A simple statistical analysis of the observations from EFF shows that 
there is apparently no correlation between profile shape and any other 
measured parameter of the cluster.  It is suggested that the prime 
mechanism for setting the profiles of these young globular clusters is the 
value of the SFE and the residual gas expulsion whose disruptive effect 
depends upon it.

It is suggested that the variety of young globular clusters observed in the 
LMC could be produced from a fairly uniform population of proto-cluster 
clouds which form stars in a relaxed distribution (similar to a Plummer model 
with length scale $1 < R_{\rm S}/$pc$ < 3$) and SFEs from 25\% to 60\% where 
the residual gas was expelled by a supershell caused by the supernovae of 
the most massive stars in the cluster.  These scale lengths correspond 
to central densities similar to those found in giant molecular clouds 
in the Galaxy (Harris \& Pudritz 1994).

The present understanding of star formation, especially in dense environments 
such as those presumably present in proto-cluster clouds, is not good enough 
to answer the question as to why SFEs should vary so much between different 
globular clusters (if, indeed, it does).  The presence and strength of 
magnetic fields within the proto-cluster cloud may have a significant 
effect upon the SFE.  The rotation of a proto-cluster cloud could, also, 
produce an effect upon the SFE.

The most uncertain assumption made is that the initial distribution of the 
clusters is similar to a Plummer model.  This assumption is made partly on the 
basis of simplicity (and the ability to use a simplification of the 
complex hydrodynamics of gas expulsion) and partly on observations of NGC 
2070 (section 2.3). 
Kennicutt \& Chu (1988) question the physical significance of the good fit 
given by star counts in NGC 2070 to an isothermal profile as stars still appear 
to be forming in NGC 2070 and most of the stars to which they fit their 
profile will, they suggest, disappear over a few 10s of Myr.  As Kennicutt \& 
Chu's survey only includes stars with $M > 10\msun$ this would 
appear to be so.  However, it seems reasonable to assume (in the absence 
of evidence to the contrary) that these stars would trace the underlying 
low mass stellar distribution.  In this case the use of a Plummer model 
would seem not to be too unreasonable a first approximation, especially if 
the high mass stars are clumped more than the low mass stars (ie. the 
mass-to-light ratio is not constant).  

The stellar population may not be virialised, and the effect that this may 
have was discussed in section 4.1.  These simulations also assumed that 
the velocity distribution is isotropic, 
which may well not be the case.  These clusters are probably rotating and 
may have other ordered internal motions.  What effects these may have is 
difficult to estimate but they are assumed to be minimal, especially when 
compared to the effects of the residual gas expulsion phase.  

Whatever the initial stellar distribution function 
it is possible to say that in a very short period of time ($<20$ Myr) these 
young globular clusters are well described by King models in their 
central regions (Chrysovergis \et 1989) which would suggest that, by this time 
they are relaxed.  A paper is in preparation in which the effects of a 
clumpy initial distribution are investigated.  Results appear to show that 
violent relaxation is able to erase substructure very rapidly, however, 
the profiles of very clumpy initial distributions are not compatible 
with observation.

The result that the three coeval clusters (NGC 2156, 2159 and 2172) within 
the quartet have profiles consistent with very similar initial proto-cluster 
conditions, but different SFEs is interesting.  These initial 
conditions are also consistent with those derived for the quartet's fourth 
member NGC 2164.  Interestingly the age difference between NGC 2164 ($\tau = 
98$ Myr) and the quartet's other members ($\tau = 65$ Myr) may suggest that 
the formation of the other three clusters may have been triggered 
by the residual gas expulsion from NGC 2164.  This is a similar scenario 
to that envisioned for the young LMC double globular cluster NGC 1850 
(Gilmozzi \et 1994).

If, as appears possible, these young LMC clusters are analogues to the 
young Galactic population in the epoch of globular cluster formation (10 to 
15 Gyr ago) they may provide information on that population.  As only 2 or 3 
of the clusters in this sample are good candidates for survival in the 
Galaxy, this may imply that the initial Galactic population was far 
larger than that observed today.  However, a full analysis of the 
possible survivability is dependent upon the slopes of the IMFs, which 
are poorly known, as well as upon full simulations of globular cluster 
evolution.

\section{Conclusions}

The conclusions of this paper may be briefly summarised as:

\begin{enumerate}

\item  The variety of young globular clusters in the LMC can be explained if 
they all formed in large proto-cluster clouds with a distribution similar 
to a Plummer model with a scale length of 1 to 3 pc (the majority with 
a scale length of around 2 pc).  The effects of residual gas expulsion 
would explain the differences in profiles between clusters if their 
star formation efficiencies varied between 25\% and 60\%.

\item The three coeval clusters from the quartet (NGCs 2156, 2159 and 2172) 
have profiles consistent with very similar initial conditions with 
$R_{\rm S} \approx 2.4$ pc if their star formation efficiencies were 
$\approx$ 40\%, 25\% and 60\% respectively.  Their age (65 Myr) is also 
consistent with their star formation being triggered by the expulsion 
of residual gas from the quartet's fourth member NGC 2164 ($\tau = 98$ Myr).

\item  If these clusters have an initial mass function slope similar to 
the Salpeter value (2.35) then 3 of these clusters (NGCs 1818, 2159 
and 2214) would appear to be good candidates for globular clusters that 
may be able to survive for a Hubble time within the Galaxy.  Within the 
LMC, NGCs 2156 and 2172 may also be able to survive for a Hubble time.

\item  If the initial mass function slope of these clusters is much lower 
than 2.35 then none of these clusters would be expected to survive for a 
significant time in the Galaxy.  Even survival for more than a few Gyr 
within the LMC would seem unlikely due to the large amounts of mass loss 
from stellar evolution.

\item  If these clusters do represent a young analogue to the Galactic 
globular clusters then the initial number of globular clusters in our 
Galaxy would have been far higher than is observed today.  The disruption 
of clusters from evaporation considered in this paper would be heightened 
by other processes such as bulge and disc shocking.  The present 
globular cluster population of the Galaxy may only represent a small 
proportion of the original population.  

\end{enumerate}

\section{Aknowledgements}

Thanks go to my supervisor R.\,J.\,Tayler, S.\,J.\,Aarseth for allowing me 
use of his nbody2 code and his useful comments as referee for this paper, 
D.\,C.\,Heggie for help with the computational side and R.\,A.\,W.\,Elson 
for discussions about the observations.  S.\,P.\,Goodwin 
is a DPhil student at the University of Sussex in receipt of a PPARC grant.

\end{document}